\documentclass[12pt,endfloat]{article}
\usepackage{color}
\usepackage{graphicx}
\usepackage{amsmath}
\usepackage{amssymb}
\usepackage{setspace}
\usepackage{fancyhdr}
\usepackage{fancybox}
\usepackage{endfloat}
\usepackage{url}

\newcount\hh
\newcount\mm
\mm=\time
\hh=\time
\divide\hh by 60
\divide\mm by 60
\multiply\mm by 60
\mm=-\mm
\advance\mm by \time
\def\hhmm{\number\hh:\ifnum\mm<10{}0\fi\number\mm}

\usepackage[round]{natbib}
\AtBeginDelayedFloats{%
 \pagestyle{empty}
 \cleardoublepage
}

\RequirePackage{titlesec}                    

\newtheorem{theorem}{Theorem}[section]

\newtheorem{lemma}[theorem]{Lemma}

\def\Ex{\mathbf{E}}

\def\bQ{\mathbb{Q}}
\def\bP{\mathbb{P}}

{\begin{center}\begin{Sbox}\begin{minipage}{\linewidth}\begin{small}\sf\color{red}(#1)}%
{\end{small}\end{minipage}\end{Sbox}\end{center}\colorbox{yellow!90!black}{\TheSbox}}
{\begin{center}\begin{Sbox}\begin{minipage}{\linewidth}\begin{small}\sf\color{black}(#1)}%
{\end{small}\end{minipage}\end{Sbox}\end{center}\colorbox{green!90!black}{\TheSbox}}
{\begin{center}\begin{Sbox}\begin{minipage}{\linewidth}\begin{small}\sf\color{black}(#1)}%
{\end{small}\end{minipage}\end{Sbox}\end{center}\colorbox{orange!90!black}{\TheSbox}}

\begin{document}

\thispagestyle{empty}

\begin{center}
{\LARGE{An automaton approach for waiting times\\[0.5cm] in DNA evolution}\\[1cm]}
\end{center}

\noindent Sarah Behrens,\\
Westf\"alische Wilhelms-Universit\"at, Institute for Evolution and Biodiversity,\\
H\"ufferstrasse 1 , 48149 M\"unster, Germany,\\
phone: +49-(0)251-83-21096,
fax: +49-(0)251-83-24668,\\
{\tt sbehrens@uni-muenster.de}\\[0.5cm]
\noindent
Cyril Nicaud,\\
LIGM, CNRS-UMR 8049, Paris-Est, France\\
phone: 33(0)16095-7550,
fax +33(0)16095-7557,\\
{\tt Cyril.Nicaud@univ-mlv.fr}\\[0.5cm]
\noindent
Pierre Nicod\`eme\footnote{corresponding author},\\
LIX, CNRS-UMR 7161, \'Ecole polytechnique,\\
91128 Palaiseau and  AMIB Team, INRIA-Saclay, France\\
phone: +33(0)16933-4112, 
fax: +33(0)16933-4049,\\
{\tt nicodeme@lix.polytechnique.fr}. \\[0.5cm]
\noindent
{\bf Running head:} Waiting times and Evolution\\[0.5cm]

\newpage

\pagestyle{plain}

\begin{abstract}
In a recent article, Behrens and Vingron (JCB 17, 12, 2010) compute waiting times for
$k$-mers to appear during DNA evolution under the assumption that the considered
$k$-mers do not occur in the initial DNA sequence, an issue arising when studying 
the evolution of regulatory DNA sequences with regard to transcription factor (TF) binding site emergence.
The mathematical analysis underlying their computation assumes that occurrences of words under interest
do not overlap. We relax here this assumption by use of an automata approach.
In an alphabet of size $4$ like the DNA alphabet,
most words have no or a low autocorrelation; therefore, globally,
our results confirm those of Behrens and Vingron.
The outcome is quite different when considering highly autocorrelated $k$-mers;
in this case, the autocorrelation pushes down the probability of occurrence of these $k$-mers at generation 1 
and, consequently, increases the waiting time for apparition of these $k$-mers up to $40\%$. 
An analysis of existing TF binding sites unveils a significant proportion of $k$-mers
exhibiting autocorrelation. Thus, our computations based on automata greatly improve the accuracy of predicting
waiting times for the emergence of TF binding sites to appear during DNA evolution.
We do the computation in the Bernoulli or M0 model; computations in
the M1 model, a Markov model of order 1, are more costly in terms of time and memory but should
produce similar results.
While Behrens and Vingron considered specifically promoters of length
$1000$,
we extend the results to promoters of any size;
we exhibit the property that the probability that a
$k$-mer occurs at generation time $1$ while being absent at time $0$ behaves
linearly with respect to the length of the promoter,
 which induces a hyperbolic behaviour of the waiting time
of
any $k$-mer with respect to the length of the promoter.
\end{abstract}

\section{Introduction}
\label{sec:intro}
The expression of genes is subject to strong regulation. The key concept of transcriptional gene regulation is the binding of proteins, so called transcription factors (TFs), to TF binding sites. These TF binding sites are typically short stretches of DNA, many of which are only around 5--8bp long (\cite{wray}). Usually, these TF binding sites are located in a region around 1000bp upstream of the gene they regulate, the so called promoter. Thus, the occurrence of particular $k$-mers in these promoter regions has a high impact on modulating transcription. 
There have been several experimental studies employing ChIP-chip or ChIP-seq technology showing that promoters are rapidly evolving regions that change over short evolutionary time scales (\cite{odom}, \cite{schmidt}, \cite{kunarso}). In a recent review, \cite{dowell} summarizes all these experimental findings and concludes that most TF binding events are species-specific and that gene regulation is a highly dynamic evolutionary process. Many of these changes in TF binding, if not necessarily all, can be explained by gains and losses of TF binding sites.

Several theoretical studies have tried to give a probabilistic explanation for the speed of changes in transcriptional gene regulation (e.g. \cite{stone}, \cite{durrett}). \cite{BehVin2010} infer how long one has to wait until a given TF binding site emerges at random in a promoter sequence. Using two different probabilistic models (a Bernoulli model denoted by M0 and a neighbor dependent model M1) and estimating evolutionary substitution rates based on multiple species promoter alignments for the three species {\it Homo sapiens}, {\it Pan troglodytes} and {\it Macaca mulatta}, they compute the expected waiting time for every $k$-mer, $k$ ranging from 5 to 10, until it appears in a human promoter. They conclude that the waiting time for a TF binding site is highly determined by its composition and that indeed TF binding sites can appear rapidly, i.e.~in a time span below the speciation time of human and chimp.

However, in their approach, \cite{BehVin2010} rely on the assumption that if a $k$-mer of interest appears more than once in a promoter sequence, it does not overlap with itself. This particularly affects the waiting times for highly autocorrelated words like e.g. {\tt AAAAA} or {\tt CTCTCTCTCT}. Using automata, we can relax this assumption and, thus, more accurately compute the expected waiting times until appearance for every $k$-mer, $k$ ranging from 5 to 10, in a promoter of length 1000bp.
This automaton approach can be applied both for models M0 and M1. However, for the ease of exposition, in this article we will focus on the Bernoulli model M0.

This article is structured as follows. In Section \ref{sec:models}, we
describe model M0, state results from \cite{BehVin2010} that we
rely on and recall how \cite{BehVin2010} have estimated model M0
parameters based on human, chimp and macaque promoter alignments. In
Section \ref{sec:auto}, we present our new approach of computing
waiting times using automata theory; we provide in this section a
web-pointer
to the program used to perform these computations.
Section \ref{sec:bioresults} 
compares the results of computing waiting times for $k$-mers to appear
in a promoter of length 1~kb according to \cite{BehVin2010} and to our
new automaton approach. For both computations, we employ the same
model parameters estimations that have been already used in
\cite{BehVin2010}; we also  explain in this section
the biological impact of our findings and show that autocorrelation
matters in the context of TF binding site emergence. 
Section \ref{sec:linear} exhibits the first order
linear behaviour of the
probability of evolution to a $k$-mer from generation time $0$ to time $1$ for specific
examples; the observed phenomena is however general, as proved
in \cite{Nicodeme2011}. We provide in this section
a web-pointer to a database containing the waiting times of all $k$-mers
for $k$ from $5$ to $10$ and for promoter lengths $n=1000$ and $n=2000$.
Section \ref{sec:conclusion} will conclude the article with some summarizing
remarks.

\section{Model M0 and expected waiting times}
\label{sec:models}
Throughout the article, we assume that promoter sequences evolve according to model M0 which has been described by \cite{BehVin2010}.
\paragraph{Model M0.}
Given an alphabet $\mathcal{A}=\{\text{A,C,G,T}\}$, let
$S(0)=(S_1(0),\dots, S_n(0))$ denote the initial promoter sequence of
length $n$ taking values in this alphabet. We assume that the letters in $S(0)$ are independent and identically distributed with $\nu(x):=\Pr(S_1(0)=x)$. 
Let the time evolution $(S(t))_{t\geq 0}$ of the promoter sequence be
governed by the $4\times4$ infinitesimal rate matrix $\bQ=(r_{\alpha,\beta})_{\alpha,\beta\in\mathcal{A}}$. According to the general reverse complement symmetric substitution model, we assume that the nucleotides evolve independently from each other and that $r_{A, T}=r_{T, A}$, $r_{C , G}=r_{G, C}$, $r_{A , C}=r_{T, G}$, $r_{C , A}=r_{G, T}$, $r_{A , G}=r_{T, C}$ and $r_{G , A}=r_{C, T}$ (see also \cite{arndt3}). Thus, there are 6 free parameters. The matrix $\bP(t)=(p_{\alpha,\beta}(t))_{\alpha,\beta\in\mathcal{A}}$ containing the transitions probabilities of $\alpha$ evolving into $\beta$ in finite time $t\geq 0$, ($\alpha,\beta\in\mathcal{A}$), can be computed by $\bP(t)=e^{t\bQ}$; see~\cite{KarTay75}, p. 150-152.

\paragraph{The expected waiting time.}
Given a binding site \begin{equation}b=(b_1,\dots,b_k)\quad\text{where } b_1,\dots,b_k\in\mathcal{A},\end{equation} the aim is to determine the expected waiting time until $b$ emerges in a promoter sequence of length $n$ provided that it does not appear in the initial promoter sequence $S(0)$.
More precisely, let
\begin{equation}
T_n=\inf\{t\in\mathbb{N} :\exists i\in\{1,\dots,n-k+1\}\text{ such that }(S_i(t),\dots,S_{i+k-1}(t))=(b_1,\dots,b_k)\}.
\end{equation}
Then, given that $\Pr(b\text{ occurs in }S(0))=0$, $T_n$ has approximately a geometric distribution with parameter 
\begin{align}
\mathfrak{p}_n&=\Pr(b\text{ occurs in generation 1}\ |\ b\text{ does not
  occur in generation 0})\\
\nonumber &= \Pr(b\in S(1) \ |\ b\not\in S(0))
\end{align}
as shown by \cite{BehVin2010}. In particular, one has
\begin{equation}
\label{eq:expected}
 \Ex(T_n)\approx\frac{1}{\mathfrak{p}_n}.
\end{equation}

\paragraph{Estimating the parameters of model M0.}
For our analyses, we used the same parameter estimations as \cite{BehVin2010}.
The estimations for $\nu(\alpha)$, $\alpha\in\mathcal{A}$, have been obtained by
determining the relative frequencies of A, C, G and T in human
promoter regions downloaded from UCSC. The substitution rates
$r_{\alpha,\beta}$ have been estimated using multiple alignments from
UCSC of chimp and macaque DNA sequences to human promoters and by
employing the Maximum likelihood based tool developed by
\cite{arndt2}. Afterwards, the transition probabilities
$p_{\alpha,\beta}(t)$ for e.g. $t=1$ generation can be easily computed
by the matrix exponential $\bP(t)=e^{t\bQ}$. Assuming a speciation
time between human and chimp of 4 Million of years and a generation
time of $y=20$ years, \cite{BehVin2010} obtain estimations for $p_{\alpha,\beta}(1)=p_{\alpha,\beta}(1\text{ generation})$ for all $\alpha,\beta\in\mathcal{A}$. Their results are summarized in Table \ref{estimations}.

\begin{table}
A) Estimations for $\nu(a)$, $a\in\mathcal{A}$:\\[0.2cm]
\begin{tabular}{|cccc|}
\hline
$\nu(A)$ & $\nu(C)$ & $\nu(G)$ & $\nu(T)$\\
\hline
0.23889 & 0.26242 & 0.25865 & 0.24004\\
 \hline
\end{tabular}\\

B) Estimations for $p_{\alpha,\beta}(1)$, $\alpha,\beta\in\mathcal{A}$:\\[0.2cm]
\begin{tabular}{|c|cccc|}
\hline
& A & C & G & T\\
\hline
A & 9.99999996e-01& 4.54999995e-09& 1.57499996e-08&3.40000002e-09\\
C & 6.14999993e-09& 9.99999996e-01& 7.14999985e-09&2.17499994e-08\\
G & 2.17499994e-08& 7.14999985e-09& 9.99999996e-01&6.14999993e-09\\
T & 3.40000002e-09& 1.57499996e-08& 4.54999995e-09& 9.99999998e-01\\
\hline
\end{tabular}
\caption{{\bf Parameter estimations.} Numbers taken from \cite{BehVin2010}, Supplementary Material S2.\label{estimations}}
\end{table}

\section{Automaton approach}
\label{sec:auto}
The aim of this section is to provide a new procedure to compute the
expected waiting time $ \Ex(T_n)$ until a TF binding site $b$ of
length $k$ emerges in a promoter sequence of length $n$ by using
Equation \eqref{eq:expected},
i.e. $\Ex(T_n)\approx\frac{1}{\mathfrak{p}_n}$. \cite{BehVin2010}
approximated $\mathfrak{p}_n=\Pr(b\text{ occurs in generation
  1}|b\text{ does not occur in generation 0})$ by applying the
inclusion-exclusion principle. However, in order to make the
computations feasible, they had to assume that $b$ cannot appear
self-overlapping which especially adulterates the actual waiting times
for autocorrelated words. Automata theory provides a natural and
compact framework to handle autocorrelations easily; in this section we present how to use basic automata algorithms in order to compute the probability $\mathfrak{p}_n$ without resorting to the assumption that $b$ occurs non-overlapping.  

\paragraph{Definitions.}
In this section, only definitions that will be used in the sequel are recalled;
more information about automata and regular languages can be found 
in \cite{HU01}.
Given a finite alphabet $\mathcal{A}$, a {\em deterministic and complete automaton} on $\mathcal{A}$ is a tuple $(Q,\delta,q_0,F)$, where $Q$ is a finite set of
{\em states}, $\delta$ is a mapping from $Q\times\mathcal{A}$ to $Q$, $q_0\in Q$ is the initial state and $F\subseteq Q$ is the {\em set of final states}. Let $\varepsilon$ denote the empty word. The mapping $\delta$ can be extended inductively to
$Q\times\mathcal{A}^*$ by setting $\delta(q,\varepsilon)=q$ for all $q\in Q$ and, for all $q\in Q$, $u\in\mathcal{A}^*$
and $\alpha\in\mathcal{A}$, $\delta(q,u\alpha) = \delta(\delta(q,u),\alpha)$. A word $u\in\mathcal{A}^*$ is {\em recognized} by the 
automaton when $\delta(q_0,u)\in F$. The {\em language recognized} by the automaton is the set of words that are recognized.

\begin{figure}
\begin{center}
\includegraphics[width=\textwidth]{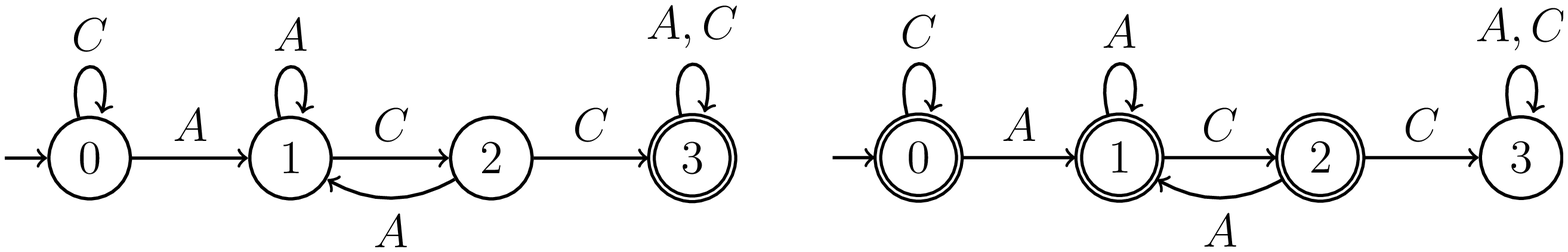}
\end{center}
\caption{\bf The automata $\mathcal{M}_{ACC}$ ($\geq 1$ occ.; on the left) and 
$\overline{\mathcal{M}}_{ACC}$ ($0$ occ.; on the right). \label{pagauto}}
\end{figure}

Since all automata considered in the sequel are deterministic and complete, we will call them ``automata'' for short. Automata
are well represented as labelled directed graphs, where the states are the vertices, and where there is an edge between
$p$ and $q$ labelled by a letter $\alpha\in\mathcal{A}$ if and only if $\delta(p,\alpha)=q$; such an edge is called a {\em transition}.
The initial state has an incoming arrow, and final states are denoted
by a double circle. See Figure~\ref{pagauto} for an example
of such a graphical representation. A word $u$ is recognized when starting at the initial state and reading $u$ from left
to right, letter by letter, and following the corresponding transition, one ends in a final state.

\paragraph{Rewording the problem.}
Consider the alphabet $\mathcal{B} = \mathcal{A}\times\mathcal{A}$. Letters of $\mathcal{B}$ are pairs
$(\alpha,\beta)$ of letters of $\mathcal{A}$, which are represented vertically by $\binom{\alpha}{\beta}$. A word $u$ 
of length $n$ on $\mathcal{B}$ is also seen as a pair of words of length $n$ over $\mathcal{A}$, and represented
vertically: if $u=(\alpha_1,\beta_1)(\alpha_2,\beta_2)\ldots(\alpha_n,\beta_n)$, we shall write $u = \binom{\alpha_1\ldots \alpha_n}{\beta_1\ldots \beta_n}$.
For any word $u=\binom{v}{w}$ of $\mathcal{B}^*$, the projections $\pi_0$ and $\pi_1$ are defined by 
$\pi_0(u) = v$ and $\pi_1(u)=w$. 

For the problems considered in this article, we have
$\mathcal{A}=\{\texttt{A,C,G,T}\}$, and a  word $u=\binom{v}{w}$ of length $n$ over $\mathcal{B}$ represents
the sequence that was initially equal to $v$ and that has evolved into $w$ at time $1$; that is, $S(0) = \pi_0(u)$ 
and $S(1)=\pi_1(u)$. The main problem can be reworded using rational expressions: for a given $b=b_1\cdots b_k$,
the fact that $b$ appear in $S(1)$ but not in $S(0)$ is exactly the condition
$\pi_1(u) \in \mathcal{A}^*b\mathcal{A}^*$ and $\pi_0(u)\notin\mathcal{A}^*b\mathcal{A}^*$. We denote by
$\mathcal{L}_b$ the set of such words and remark that $\mathcal{L}_b$
is
a rational language.

\paragraph{Construction of the automaton.}
The smallest automaton  $\mathcal{M}_b$ that recognizes the language $\mathcal{A}^*b\mathcal{A}^*$ 
can be built using the classical Knuth-Morris-Pratt construction (see~\cite{CroRyt94}, chapter 7). 
This requires for any $k$-mer $O(k)$ time and space, and the produced automaton 
$\mathcal{M}_b=(\{0,\ldots,k\},\delta_b,0,\{k\})$ has exactly $k+1$ states.

\begin{figure}
\begin{center}
\includegraphics[width=\textwidth1]{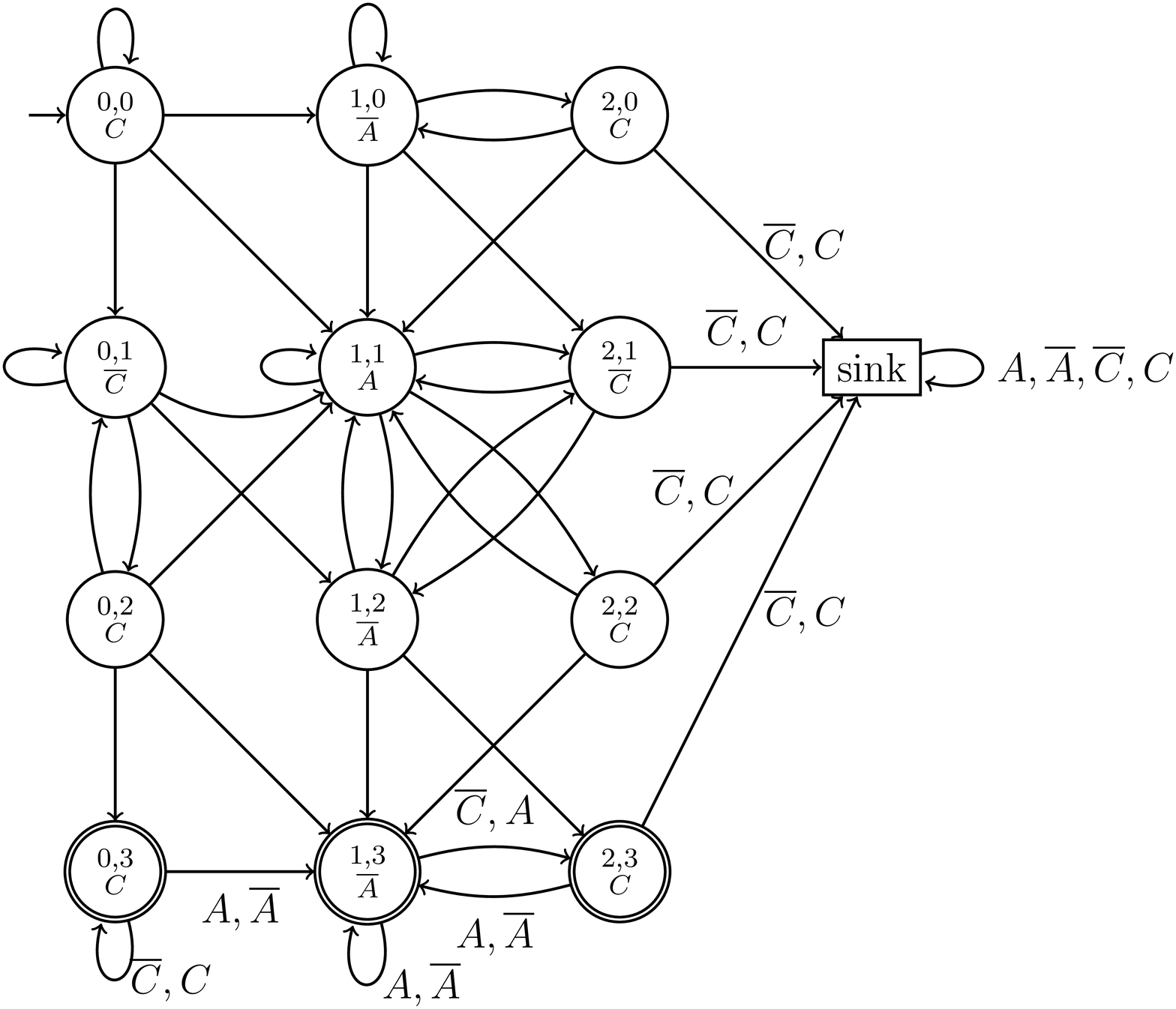}
\end{center}
\caption{{\bf The automaton $\mathcal{N}'_{ACC}$.} For the automaton to be readable, we use the notations
        $A =\binom{A}{A}$, $\overline{A}=\binom{A}{C}$, $C=\binom{C}{C}$ and $\overline{C}=\binom{C}{A}$. When the label of a transition is not given, 
        it is by default set to the letter at the bottom of its ending state.\label{pagprod}}
\end{figure}

The language $\mathcal{A}^*\setminus \mathcal{A}^*b\mathcal{A}^*$ is the complement of the previous one, and is therefore
recognized by the automaton $\overline{\mathcal{M}}_b = (\{0,\ldots,k\},\delta_b,0,\{0,\ldots,k-1\})$,
which has the same underlying graph as $\mathcal{M}_b$ and whose set of final states is the complement of $\mathcal{M}_b$'s one.
For the examples given in this section, we use a smaller alphabet $\mathcal{A}=\{A,C\}$ and the $k$-mer is always $b=ACC$, (hence $k=3$). 
The two automata are depicted in Figure~\ref{pagauto}.
To fully describe the language $\mathcal{L}_b$, we use the classical product automaton construction, tuned to fit our
needs. Define the automaton $\mathcal{N}_b=(Q,\delta,q_0,F)$ as follows:
\begin{itemize}
\item The set of states is $Q=\{0,\ldots,k\}\times\{0,\ldots,k\}$. The states of $\mathcal{N}_b$ are therefore pairs
$(p,q)$, where intuitively $p$ lies in $\overline{\mathcal{M}}_b$ and $q$ lies in $\mathcal{M}_b$.
\item The initial state is $q_0=(0,0)$.
\item The transition mapping $\delta$ is defined for every $(p,q)\in Q$ and every $(\alpha,\beta)\in\mathcal{B}$ 
by $\delta((p,q),(\alpha,\beta))=(\delta_b(p,\alpha),\delta_b(q,\beta))$. The idea is to read $\pi_0(u)$ in $\overline{\mathcal{M}_b}$
on the first coordinate, and $\pi_1(u)$ in $\mathcal{M}_b$   on the second coordinate.
\item A state $(p,q)$ is final if and only if both $p$ and $q$ are final in their respective automata, that is,
$F = \{0,\ldots,k-1\}\times\{k\}$.
\end{itemize}
The proof of the following lemma follows directly from the construction of $\mathcal{N}_b$:
\begin{lemma}\label{lm Nb}
The automaton $\mathcal{N}_b$ recognizes the language $\mathcal{L}_b$.
\end{lemma}
Looking closer at the automaton one can make the following observations: while reading a word $u$ of $\mathcal{B}^*$
in $\mathcal{N}_b$, if one reaches a state of the form $(p,k)$ at some point, for some $p\in\{0,\ldots,k\}$, then all 
the remaining states on the path labelled by $u$ are also of the form $(q,k)$, for some $q\in\{0,\ldots,k\}$. This is
because $\delta_b(k,\alpha)=k$ for every $\alpha\in\mathcal{A}$. Since this state is not final, this means that whenever
the second coordinate is $k$ at some point, the word is not recognized because $\pi_0(u)$ contains $b$. We can therefore simplify
the automaton $\mathcal{N}_b$ by merging all the states of the form $(p,k)$ into a single state, which we name {\em sink}. 
Let $\mathcal{N}'_b=(Q',\delta',q_0',F')$ denote this new automaton, which has $k^2+k+1$ states. Lemma~\ref{lm Nb prime} below
states that all the information we need is contained in
$\mathcal{N}'_b$.
See an example of this automaton in Figure \ref{pagprod}.
\begin{lemma}\label{lm Nb prime}
Let $u$ be a word in $\mathcal{B}^*$, and let $q_u$ be the state reached after reading $u$ in $\mathcal{N}_b'$ from its initial
state. The words $u$ can be classified as follows:
\begin{itemize}
\item if $q_u\in F'$ then $\pi_0(u)$ does not contains $b$ but $\pi_1(u)$ does (this is a success in our settings);
\item if $q_u$ is the sink state then $\pi_0(u)$ contains $b$ (this is contradictory in our settings);
\item if $q_u\notin F'$ and $q_u$ is not the sink state, then neither $\pi_0(u)$ nor $\pi_1(u)$ contains $b$ (this is a failure in our settings).
\end{itemize} 
\end{lemma}

\paragraph{From automata to probabilities.}
The automaton $\mathcal{N}'_b$ is readily transformed into a Markov chain, by changing the label of any transition
$q\xrightarrow{a}q'$, where $a=\binom{\alpha}{\beta}\in \mathcal{B}$, into the probability $\nu(\alpha)\times p_{\alpha,\beta}(1)$.
If there are several transitions from $q$ to $q'$, the edge is labelled by the sum of the associated probabilities. Let
$\mathcal{C}_b$ denote this Markov chain. The random variable $Q_n$ associated to the state reached after reading a random word of size $n$ under the M0 model
is formally defined by:
\begin{equation}
\forall q\in Q',\  
\Pr\left(Q_n=q\right) = \sum_{\substack{u=\binom{v}{w}\in\mathcal{B}^n\\ \delta'(q'_0,u)=q}} \nu(v)\times p_{v\rightarrow w}(1).
\end{equation}
Then, if $\mathbb{P}_b$ is the transition matrix of $\mathcal{C}_b$ and if $V_{q}$ is the probability vector with $1$ on
position $q\in Q'$ and $0$ elsewhere, the random state $Q_n$ reached
from the initial state after $n$ steps verifies
\begin{equation}\label{matrix formula}
\forall q\in Q',\  
\Pr\left(Q_n=q\right) = V_{q'_0}^t\times \mathbb{P}_b^n\times V_{q}.
\end{equation}
From this and by Lemma~\ref{lm Nb prime} we can compute all the needed probabilities~:
\begin{align}
\label{eq:S1S0}
\Pr\Big(S(1)\in \mathcal{A}^*b\mathcal{A}^*\mid S(0)\notin \mathcal{A}^*b\mathcal{A}^*\Big) & = \frac{\Pr(S(1)\in \mathcal{A}^*b\mathcal{A}^*\text{ and } S(0)\notin \mathcal{A}^*b\mathcal{A}^*)}{\Pr(S(0)\notin \mathcal{A}^*b\mathcal{A}^*)} \\
& = \frac{\Pr(Q_n\in F')}{\Pr(Q_n=\text{sink})}\\
& = \frac{\sum_{q\in F'}V_{q'_0}^t\times \mathbb{P}_b^n\times V_{q}}{V_{q'_0}^t\times \mathbb{P}_b^n\times V_{\text{sink}}}
\end{align}
We therefore get our main result.
\begin{theorem}\label{th main}
Let $b\in\mathcal{A}^k$ and $\mathcal{N}'_b=(Q',\delta',q'_0,F')$ be
its automaton, with associated matrix $\mathbb{P}_b$. The probability $\mathfrak{p}_n$ that a sequence
of length $n$ contains $b$ at time $1$ given that it does not contains $b$ at time $0$ is
exactly
\[\mathfrak{p}_n=
\Pr\Big(S(1)\in \mathcal{A}^*b\mathcal{A}^*\mid S(0)\notin \mathcal{A}^*b\mathcal{A}^*\Big)  = \frac{V_{q'_0}^t\times \mathbb{P}_b^n\times \left(\sum_{q\in F'} V_{q}\right)}{V_{q'_0}^t\times \mathbb{P}_b^n\times V_{\text{sink}}}.
\]
\end{theorem}
Applying Theorem \ref{th main} and Equation \eqref{eq:expected}, we obtain that the expected waiting time $\Ex(T_n)\approx\frac{1}{\mathfrak{p}_n}$ until a binding site $b$ of length $k$ appears in a promoter of length $n$ can be approximated by
\begin{equation}
\label{eq:automata}
\Ex(T_n)\approx\frac{1}{\mathfrak{p}_n}=\frac{1}{\Pr\Big(S(1)\in \mathcal{A}^*b\mathcal{A}^*\mid S(0)\notin \mathcal{A}^*b\mathcal{A}^*\Big)}=\frac{V_{q'_0}^t\times \mathbb{P}_b^n\times V_{\text{sink}}}{V_{q'_0}^t\times \mathbb{P}_b^n\times \left(\sum_{q\in F'} V_{q}\right)}.
\end{equation}

\paragraph{Complexity.} 
The automaton $\mathcal{N}'_b$, and the associated Markov chain $\mathcal{C}_b$ can be built in time and space $O(|\mathcal{A}|^2k^2)$.
Once done, the whole calculation reduces to the computation of the row vector $V_{q'_0}^t\times \mathbb{P}_b^n$, which can be done iteratively using the simple relation
\[
V_{q'_0}^t\times \mathbb{P}_b^{i+1} = \underbrace{\left(V_{q'_0}^t\times \mathbb{P}_b^{i}\right)}_{\text{row vector}}\times \mathbb{P}_b.
\]
Hence this consists of $n$ products of a vector by a matrix. Moreover, 
this matrix is a square matrix of dimension $k^2+k+1$, which is sparse
since it has exactly $(k^2+k+1)|\mathcal{A}|^2$ non-zero values. Therefore,
the probability of Theorem~\ref{th main} can be computed in time
$O(n\times k^2\times |\mathcal{A}|^2)$, using $O(|\mathcal{A}|^2k^2)$ space.

\paragraph{Web access to the code.}\footnotesize \hskip-2ex  URL
 \scriptsize\mbox{\url{http://www.lix.polytechnique.fr/Labo/Pierre.Nicodeme/BNN/kmer.c}}
\normalsize
 provides the \texttt{C} code used in this section.

\nocite{GuiOdl81a,GuiOdl81b,GoJa83,Lot05,FlajoletSedgewick2009,BehVin2010}

\section{Biological results}
\label{sec:bioresults}
Applying  Equation \eqref{eq:automata} for obtaining the automaton
results 
and using Theorem 1 from \cite{BehVin2010}, we computed the expected
waiting time $\Ex(T_{1000})$
of all $k$-mers in the M0 model for $k$ from $5$ to $10$ to appear in a promoter sequence of length 1000~bp.
The parameters of model M0 have been estimated as described in Section \ref{sec:models} and are depicted in Table \ref{estimations}.

\begin{figure}
\begin{center}
\includegraphics[width=\textwidth]{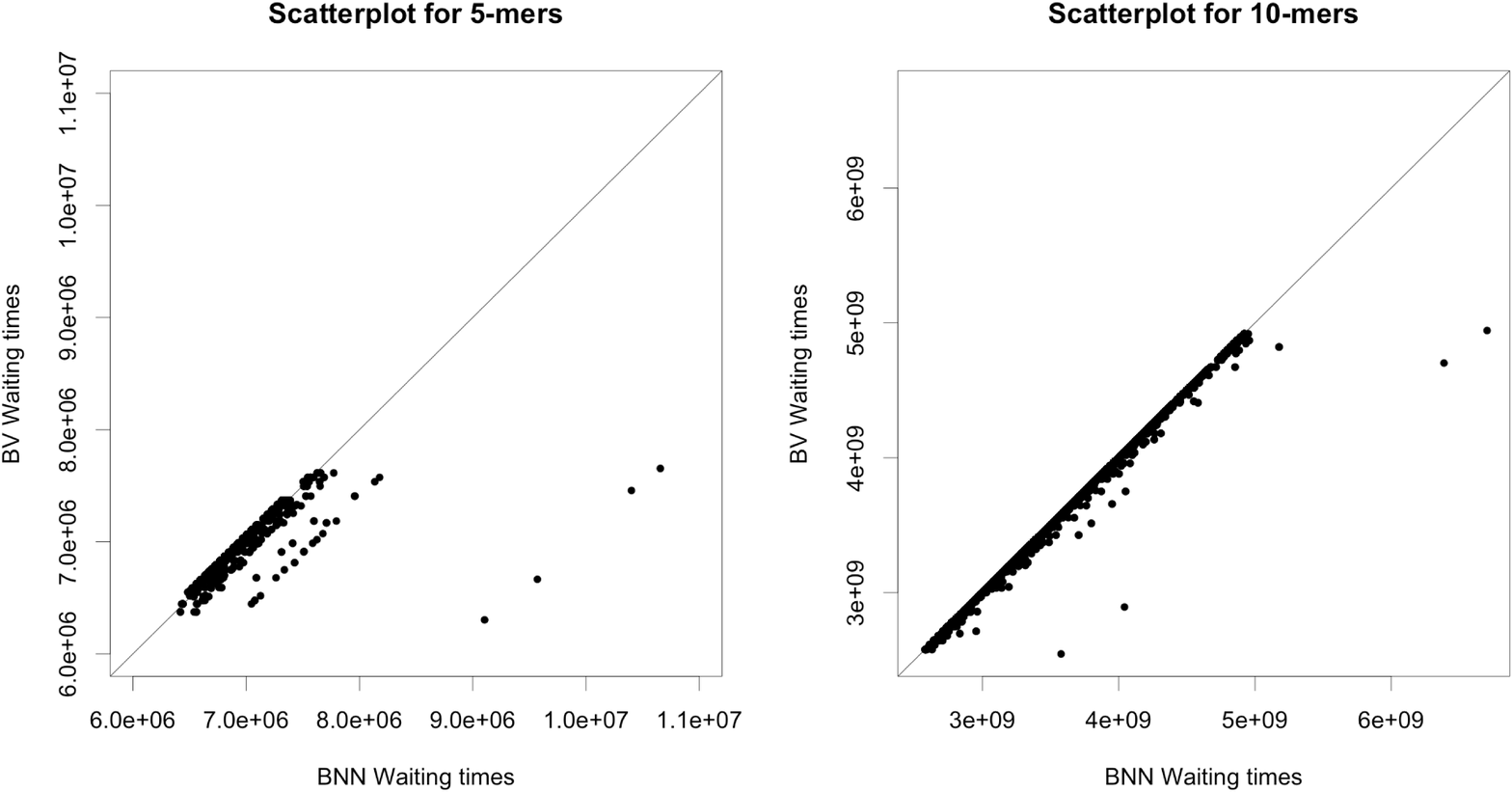}
\end{center}
\caption{\label{fig:scatter} {\bf Overall comparisons of waiting times
of \cite{BehVin2010} (BV) versus the automata method (BNN) for $5$-
and $10$-mers.}}
\end{figure}

Figure~\ref{fig:scatter} provides an overall comparison of the waiting
time computed by automata with respect to the previous computations
of \cite{BehVin2010} for $k=5$ and $k=10$. 
As can be observed in this scatterplot, the computed waiting times based on the automaton approach globally confirm the results of \cite{BehVin2010}. However, there are some outliers exhibiting longer waiting times than predicted by \cite{BehVin2010}.
The four most extreme outliers that deviate from the bisecting line correspond to \texttt{AAAAA}, \texttt{TTTTT}, \texttt{CCCCC}, \texttt{GGGGG} and to \texttt{AAAAAAAAAA}, \texttt{CCCCCCCCCC}, \texttt{GGGGGGGGGG}, \texttt{TTTTTTTTTT} respectively.
Other outliers are $k$-mers like e.g.~\texttt{CGCGC}, \texttt{TCTCT} and \texttt{CGCGCGCGCG}, \texttt{TCTCTCTCTC}.
Tables~\ref{tab:corrank5}, \ref{tab:corrank7} and
\ref{tab:corrank10} show all $5$-, $7$- and $10$-mers for which $\frac{\Ex_{\operatorname{BNN}}(T_{1000})}{\Ex_{\operatorname{BV}}(T_{1000})}>1.05$ where $\Ex_{\operatorname{BV}}(T_{1000})$ denotes the expected waiting time according to \cite{BehVin2010} and $\Ex_{\operatorname{BNN}}(T_{1000})$ according to our automaton approach, i.e.~$k$-mers with significantly longer waiting times than predicted by \cite{BehVin2010}. 
\begin{table}
\begin{tabular}{|c||rr||rr||c|}\hline
  &  \multicolumn{2}{|c||}{BNN} & \multicolumn{2}{|c||}{BV} &  \\
  &$\Ex_{\operatorname{BNN}}(T_{1000})/10^6$& Rank & $\Ex_{\operatorname{BV}}(T_{1000})/10^6$ & Rank & $\frac{\Ex_{\operatorname{BNN}}(T_{1000})}{\Ex_{\operatorname{BV}}(T_{1000})}$ \\\hline
{\tt CCCCC} &  9.105 &      1021 &  6.304 &        1 & 1.44\\
{\tt GGGGG} &  9.570 &      1022 &  6.666 &      142 & 1.44\\
{\tt TTTTT} & 10.401 &      1023 &  7.457 &      993 & 1.39\\
{\tt AAAAA} & 10.656 &      1024 &  7.654 &     1024 & 1.39\\
{\tt CGCGC} &  7.047 &       699 &  6.446 &       11 & 1.09\\
{\tt TCCCC} &  7.076 &       737 &  6.477 &       17 & 1.09\\
{\tt CCCCT} &  7.076 &       738 &  6.477 &       21 & 1.09\\
{\tt GCGCG} &  7.127 &       787 &  6.518 &       31 & 1.09\\
{\tt CTCTC} &  7.263 &       883 &  6.679 &      148 & 1.09\\
{\tt CACAC} &  7.337 &       945 &  6.750 &      217 & 1.09\\
{\tt GGGGA} &  7.428 &       971 &  6.814 &      318 & 1.09\\
{\tt AGGGG} &  7.428 &       972 &  6.814 &      322 & 1.09\\
{\tt TCTCT} &  7.508 &       978 &  6.910 &      477 & 1.09\\
{\tt GTGTG} &  7.511 &       981 &  6.914 &      486 & 1.09\\
{\tt GAGAG} &  7.587 &       997 &  6.987 &      573 & 1.09\\
{\tt ACACA} &  7.625 &      1002 &  7.019 &      605 & 1.09\\
{\tt TGTGT} &  7.677 &      1010 &  7.073 &      735 & 1.09\\
{\tt AGAGA} &  7.796 &      1016 &  7.185 &      833 & 1.09\\
{\tt TTTTC} &  7.710 &      1013 &  7.169 &      823 & 1.08\\
{\tt CTTTT} &  7.710 &      1014 &  7.169 &      827 & 1.08\\
{\tt TATAT} &  8.135 &      1019 &  7.535 &     1003 & 1.08\\
{\tt ATATA} &  8.178 &      1020 &  7.575 &     1014 & 1.08\\
{\tt GAAAA} &  7.959 &      1017 &  7.407 &      988 & 1.07\\
{\tt AAAAG} &  7.959 &      1018 &  7.407 &      992 & 1.07\\
{\tt TTCCC} &  7.090 &       751 &  6.679 &      144 & 1.06\\
{\tt CCCTT} &  7.090 &       752 &  6.679 &      152 & 1.06\\
{\tt TTTCC} &  7.312 &       924 &  6.910 &      473 & 1.06\\
{\tt CCTTT} &  7.312 &       925 &  6.910 &      481 & 1.06\\
{\tt GGGAA} &  7.411 &       966 &  6.987 &      574 & 1.06\\
{\tt AAGGG} &  7.411 &       967 &  6.987 &      582 & 1.06\\
{\tt GGAAA} &  7.599 &      1000 &  7.185 &      828 & 1.06\\
{\tt AAAGG} &  7.599 &      1001 &  7.185 &      837 & 1.06\\
\hline\end{tabular}
\caption{\label{tab:corrank5} {\bf Expected waiting times
    (generations)
for 5-mers in model M0 with $\frac{\Ex_{\operatorname{BNN}}(T_{1000})}{\Ex_{\operatorname{BV}}(T_{1000})}>1.05$.} $\Ex_{\operatorname{BV}}(T_{1000})$ denotes the expected waiting time according to \cite{BehVin2010} (BV) and $\Ex_{\operatorname{BNN}}(T_{1000})$ according to our automaton approach (BNN). Ranks refer to $5$-mers sorted by their waiting time of appearance according to the two different procedures BV and BNN; rank 1 is assigned to the fastest evolving 5-mer, rank 1024 (=$4^5$) to the slowest emerging 5-mer.}
\end{table}

\begin{table}
\begin{tabular}{|c||rr||rr||c|}\hline
  &  \multicolumn{2}{|c||}{BNN} & \multicolumn{2}{|c||}{BV} &  \\
  &$\Ex_{\operatorname{BNN}}(T_{1000})/10^6$& Rank & $\Ex_{\operatorname{BV}}(T_{1000})/10^6$ & Rank & $\frac{\Ex_{\operatorname{BNN}}(T_{1000})}{\Ex_{\operatorname{BV}}(T_{1000})}$ \\\hline
{\tt CCCCCCC} & 93.457 &     16257 & 65.518 &        1 & 1.43\\
{\tt GGGGGGG} & 101.108 &     16380 & 71.312 &      576 & 1.42\\
{\tt TTTTTTT} & 127.536 &     16383 & 92.632 &    16257 & 1.38\\
{\tt AAAAAAA} & 131.923 &     16384 & 95.990 &    16384 & 1.37\\
{\tt CGCGCGC} & 74.347 &      2328 & 67.939 &       50 & 1.09\\
{\tt GCGCGCG} & 75.250 &      3170 & 68.766 &       86 & 1.09\\
{\tt CTCTCTC} & 81.865 &     10928 & 75.280 &     3235 & 1.09\\
{\tt CACACAC} & 83.101 &     12466 & 76.448 &     4042 & 1.09\\
{\tt GTGTGTG} & 85.914 &     14531 & 79.102 &     7786 & 1.09\\
{\tt TCTCTCT} & 85.978 &     14535 & 79.117 &     7829 & 1.09\\
{\tt GAGAGAG} & 87.211 &     15312 & 80.329 &     8656 & 1.09\\
{\tt ACACACA} & 87.721 &     15337 & 80.754 &     9267 & 1.09\\
{\tt TGTGTGT} & 89.145 &     15620 & 82.131 &    11616 & 1.09\\
{\tt TATATAT} & 101.469 &     16381 & 94.057 &    16304 & 1.08\\
{\tt ATATATA} & 101.988 &     16382 & 94.536 &    16338 & 1.08\\
{\tt AGAGAGA} & 90.953 &     16191 & 83.829 &    12794 & 1.08\\
{\tt TCCCCCC} & 73.461 &      1495 & 68.495 &       65 & 1.07\\
{\tt CCCCCCT} & 73.461 &      1496 & 68.495 &       71 & 1.07\\
{\tt GGGGGGA} & 79.292 &      7867 & 74.080 &     2158 & 1.07\\
{\tt AGGGGGG} & 79.292 &      7868 & 74.080 &     2153 & 1.07\\
{\tt TTTTTTC} & 92.782 &     16249 & 87.773 &    15367 & 1.06\\
{\tt CTTTTTT} & 92.782 &     16250 & 87.773 &    15366 & 1.06\\
{\tt GAAAAAA} & 96.810 &     16376 & 91.645 &    16255 & 1.06\\
{\tt AAAAAAG} & 96.810 &     16377 & 91.645 &    16254 & 1.06\\
\hline\end{tabular}
\caption{\label{tab:corrank7} {\bf Expected waiting times
    (generations)
for 7-mers in model M0 with $\frac{\Ex_{\operatorname{BNN}}(T_{1000})}{\Ex_{\operatorname{BV}}(T_{1000})}>1.05$.} $\Ex_{\operatorname{BV}}(T_{1000})$ denotes the expected waiting time according to \cite{BehVin2010} (BV) and $\Ex_{\operatorname{BNN}}(T_{1000})$ according to our automaton approach (BNN). Ranks refer to $7$-mers sorted by their waiting time of appearance according to the two different procedures BV and BNN; rank 1 is assigned to the fastest evolving 7-mer, rank 16384 (=$4^7$) to the slowest emerging 7-mer.}
\end{table}
\begin{table}
\begin{tabular}{|c||rr||rr||c|}\hline
  &  \multicolumn{2}{|c||}{BNN} & \multicolumn{2}{|c||}{BV} &  \\
  &$\Ex_{\operatorname{BNN}}(T_{1000})/10^6$& Rank & $\Ex_{\operatorname{BV}}(T_{1000})/10^6$ & Rank & $\frac{\Ex_{\operatorname{BNN}}(T_{1000})}{\Ex_{\operatorname{BV}}(T_{1000})}$ \\\hline
{\tt CCCCCCCCCC} & 3577.003 &    511668 & 2545.561 &        1 & 1.41\\
{\tt GGGGGGGGGG} & 4042.505 &    937454 & 2893.573 &     8844 & 1.40\\
{\tt TTTTTTTTTT} & 6387.187 &   1048575 & 4702.438 &  1047553 & 1.36\\
{\tt AAAAAAAAAA} & 6703.254 &   1048576 & 4943.605 &  1048576 & 1.36\\
{\tt GCGCGCGCGC} & 2953.939 &     16095 & 2713.901 &      443 & 1.09\\
{\tt CGCGCGCGCG} & 2953.939 &     16096 & 2713.901 &      523 & 1.09\\
{\tt TCTCTCTCTC} & 3706.263 &    658915 & 3426.738 &   337146 & 1.08\\
{\tt CTCTCTCTCT} & 3706.263 &    658916 & 3426.738 &   337202 & 1.08\\
{\tt CACACACACA} & 3799.148 &    773143 & 3513.991 &   421031 & 1.08\\
{\tt ACACACACAC} & 3799.148 &    773144 & 3513.991 &   421142 & 1.08\\
{\tt TGTGTGTGTG} & 3951.253 &    876168 & 3657.531 &   625393 & 1.08\\
{\tt GTGTGTGTGT} & 3951.253 &    876169 & 3657.531 &   625471 & 1.08\\
{\tt GAGAGAGAGA} & 4050.273 &    950059 & 3750.629 &   702887 & 1.08\\
{\tt AGAGAGAGAG} & 4050.273 &    950060 & 3750.629 &   703066 & 1.08\\
{\tt TATATATATA} & 5176.970 &   1048573 & 4821.512 &  1048005 & 1.07\\
{\tt ATATATATAT} & 5176.970 &   1048574 & 4821.512 &  1048120 & 1.07\\
\hline\end{tabular}
\caption{\label{tab:corrank10} {\bf Expected waiting times
    (generations) for 10-mers
    in model M0 with
    $\frac{\Ex_{\operatorname{BNN}}(T_{1000})}{\Ex_{\operatorname{BV}}(T_{1000})}>1.05$.}
  $\Ex_{\operatorname{BV}}(T_{1000})$ denotes the expected waiting
  time according to \cite{BehVin2010} (BV) and
  $\Ex_{\operatorname{BNN}}(T_{1000})$ according to our automaton
  approach (BNN). Ranks refer to $10$-mers sorted by their waiting
  time of appearance according to the two different procedures BV and
  BNN; rank 1 is assigned to the fastest evolving 10-mer, rank 1048576
  (=$4^{10}$) to the slowest emerging 10-mer. }
\end{table}
We use in the following the million of generations (in short Mgen) as
unit of time, where a generation is 20 years.
The discrepancy between the two procedures can attain up to around
40\%, e.g. \texttt{CCCCC} has a discrepancy of 44\% with
$\Ex_{\operatorname{BNN}}(T_{1000})=9.105\text{~Mgen}$ and
$\Ex_{\operatorname{BV}}(T_{1000})=6.304\text{~Mgen}$,
\texttt{CCCCCCC} a discrepancy of 43\% with
$\Ex_{\operatorname{BNN}}(T_{1000})=93.457\text{~Mgen}$ and
$\Ex_{\operatorname{BV}}(T_{1000})=65.518\text{~Mgen}$, and
\texttt{CCCCCCCCCC} has a discrepancy of 41\% with
$\Ex_{\operatorname{BNN}}(T_{1000})$ $=3577.003\text{~Mgen}$ and $\Ex_{\operatorname{BV}}(T_{1000})=2545.561\text{~Mgen}$. Strikingly, most of the $k$-mers with significant discrepancy feature a high autocorrelation, i.e. they can appear overlapping in so called clumps. For example, the 5-mer \texttt{CCCCC} could appear twice in the clump \texttt{CCCCCC} (at positions 1 and 2), \texttt{CGCGC} could appear three times in the clump \texttt{CGCGCGCGC} (at positions 1, 3 and 5). In order to distinguish between different levels of autocorrelation of $k$-mers, let
\[
\mathcal{P}(b):=\{p\in\{1,\dots,k-1\}:b_i=b_{i+p}\text{ for all }i=1,\dots,k-p\}
\]
denote the set of periods of a $k$-mer $b=(b_1,\dots,b_k)$. A $k$-mer $b$ is called non-periodic or non-autocorrelated if and only if $\mathcal{P}(b)=\emptyset$. Furthermore, for a periodic $k$-mer $b$ let $p_0(b)$ denote its minimal period. For example, $p_0(\texttt{CCCCC})=1$, $p_0(\texttt{CGCGC})=2$, $p_0(\texttt{CGACG})=3$ and $p_0(\texttt{CGATC})=4$. We then call a word $p$-periodic if and only if its minimal period is $p$. As can be observed in Tables~\ref{tab:corrank5}, \ref{tab:corrank7} and \ref{tab:corrank10}, half of the 5-mers, two-thirds of the 7-mers and all of the 10-mers with $\frac{\Ex_{\operatorname{BNN}}(T_{1000})}{\Ex_{\operatorname{BV}}(T_{1000})}>1.05$ are either 1- or 2-periodic, i.e. show a high degree of autocorrelation.

\cite{BehVin2010} already investigated the speed of TF binding site emergence and its biological implications for the evolution of transcriptional regulation in detail and we do not want to elaborate on this again. However, in line with \cite{BehVin2010}, we want to emphasize that the speed of TF binding site emergence is primarily influenced by its nucleotide composition. The goal in the following will be to investigate the impact of autocorrelation regarding TF binding sites. More precisely, we want to answer the question: Do existing TF binding sites show significant autocorrelation or can this aspect be neglected when studying the speed of TF binding site emergence?

\begin{figure}
\includegraphics[height=\textwidth,angle=270]{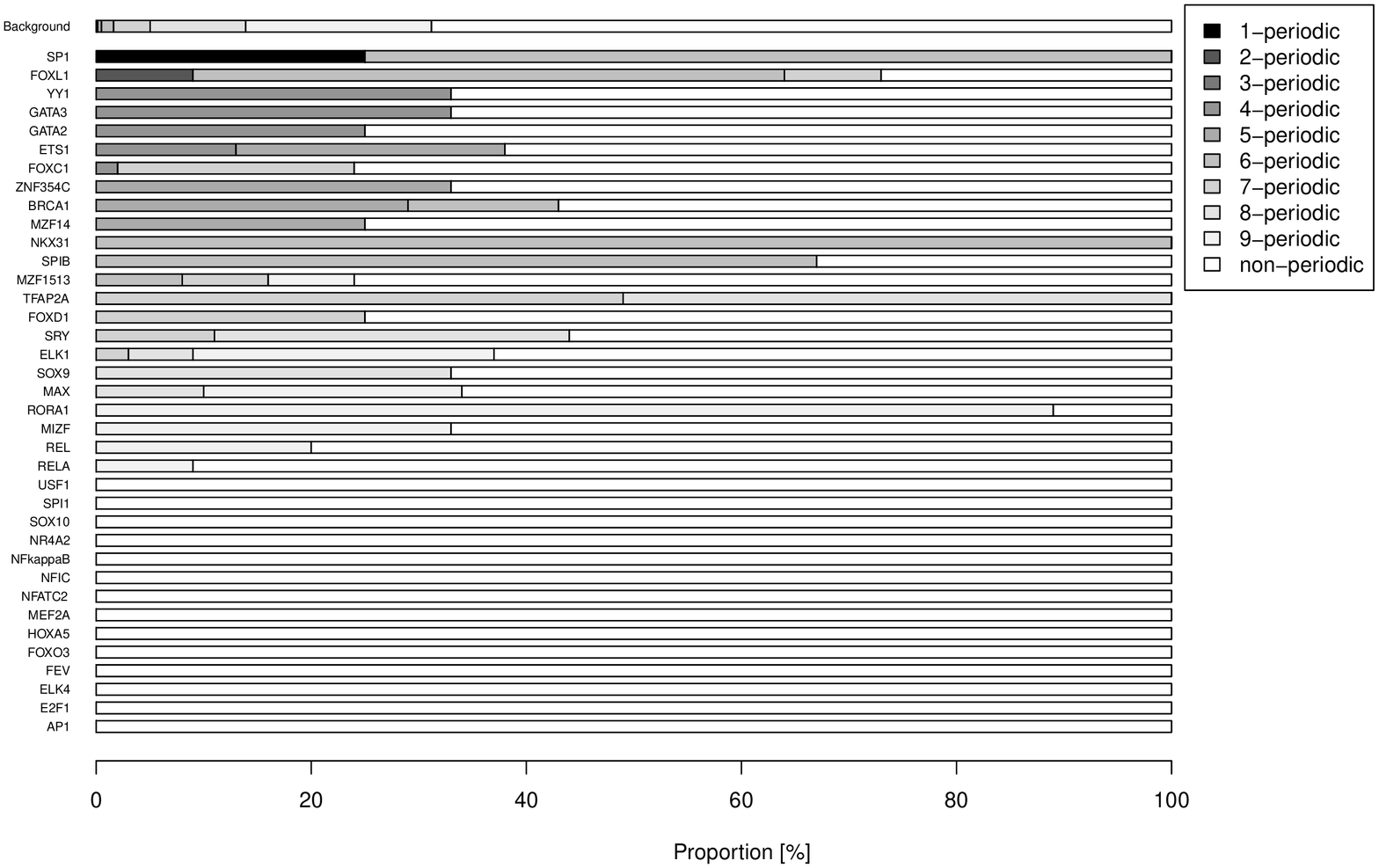}
\caption{{\bf Barplot of the degree of autocorrelation of JASPAR TF binding sites.} For every of the 37 JASPAR TFs each given by a set of $k$-mers, the proportion of $p$-periodic and non-periodic $k$-mers in this set was calculated, $p$ ranging from 1 to $k-1$. Additionally, the same proportions were computed for all possible $k$-mers, $k$ ranging from 5 to 10 ("Background").\label{fig:jaspar}}
\end{figure}

To investigate this, starting from the JASPAR CORE database for vertebrates Version 4 (\cite{jaspar}), we extracted all the human TF binding sites of length $k$, $5\leq k\leq 10$, ending up with a set of 37 position count matrices (PCMs) for the 37 different TFs in analogy to \cite{BehVin2010}. 
In order to make these PCMS accessible for our framework based on $k$-mers, we converted a PCM into a set of $k$-mers by setting a threshold of 0.95 of the maximal PCM score and extracted all $k$-mers with a score above this threshold. For example, the PCM 
\[
\begin{array}{c}
\text{A}\\
\text{C}\\
\text{G}\\
\text{T}\\
\end{array}
\left(
\begin{array}{cccccccccc}
0&0&0&4&2&0&1&0&6&3\\
32&30&35&27&5&28&31&24&25&26\\
1&1&0&0&15&1&0&3&0&3\\
2&4&0&4&13&6&3&8&4&3 
\end{array}
\right)
\]
of the TF SP1 is then translated into the following set of 10-mers: $\{\texttt{CCCCACCCCC}$, $\texttt{CCCCCCCCCC}$, $\texttt{CCCCGCCCCC}$, $\texttt{CCCCTCCCCC}\}$. Applying this procedure, in total we obtain 372 different JASPAR $k$-mers, $5\leq k\leq10$, for the 37 different human TFs. 
We then screened all JASPAR $k$-mers for 1-periodicity, 2-periodicity,..., $(k-1)$-periodicity. To evaluate the degree of autocorrelation of a given JASPAR TF given by its set of $k$-mers, we then computed the proportion of 1-periodic, 2-periodic,..., $(k-1)$-periodic and of non-periodic $k$-mers in this set. The results are depicted in Figure \ref{fig:jaspar}. 
As can be seen, some TFs like SP1, FOXL1, YY1, GATA3, GATA2 and ETS1 exhibit a high autocorrelation while 14 of the 37 TFs show no autocorrelation at all (USF1, SPI1,..., AP1). In order to test whether autocorrelated $k$-mers are enriched among JASPAR TF binding sites, as a background we screened all possible $k$-mers, i.e. all $b=(b_1,\dots,b_k)\in\mathcal{A}^k$, $\mathcal{A}=\{\text{A,C,G,T}\}$, $k$ ranging from 5 to 10, for autocorrelation in the same way as JASPAR $k$-mers. The resulting proportions of periodic and non-periodic words of this background are also depicted in Figure \ref{fig:jaspar}. In total, among the JASPAR $k$-mers, there are 168 autocorrelated words (i.e. words that are $p$-periodic for one $p\in\{1,\dots,k-1\}$) and 204 non-autocorrelated words. The background set contains 435,828 autocorrelated and 961,932 non-autocorrelated $k$-mers. Performing Fisher's Exact Test for Count Data with the alternative "greater", we obtain a $p$-value of 1.119e-08. We can thus conclude that autocorrelated words are significantly enriched among JASPAR $k$-mers. Consequently, existing TF binding sites indeed feature a significant proportion of autocorrelation.

\section{Linear behaviour of $\mathfrak{P}_n$}
\label{sec:linear}
\begin{figure}[!h]
\ \vspace*{6cm}
\setlength{\unitlength}{1.1mm}
\centering\begin{picture}(60,0)(0,20)

\put(-30,-28) {\includegraphics[height=5.6cm,width=5.6cm]{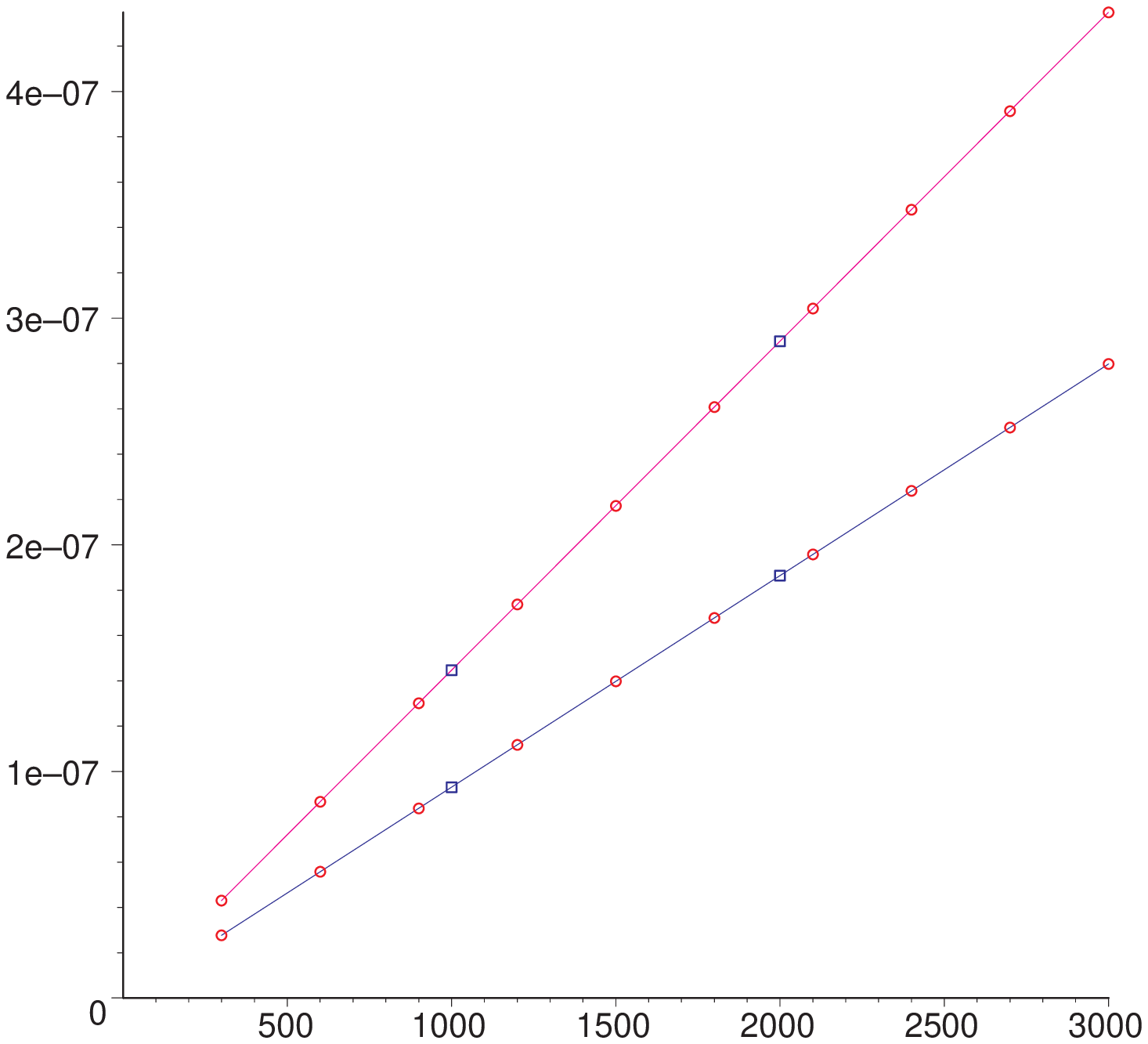}}
\put(40,-31)  {\includegraphics[height=5.6cm,width=5.6cm]{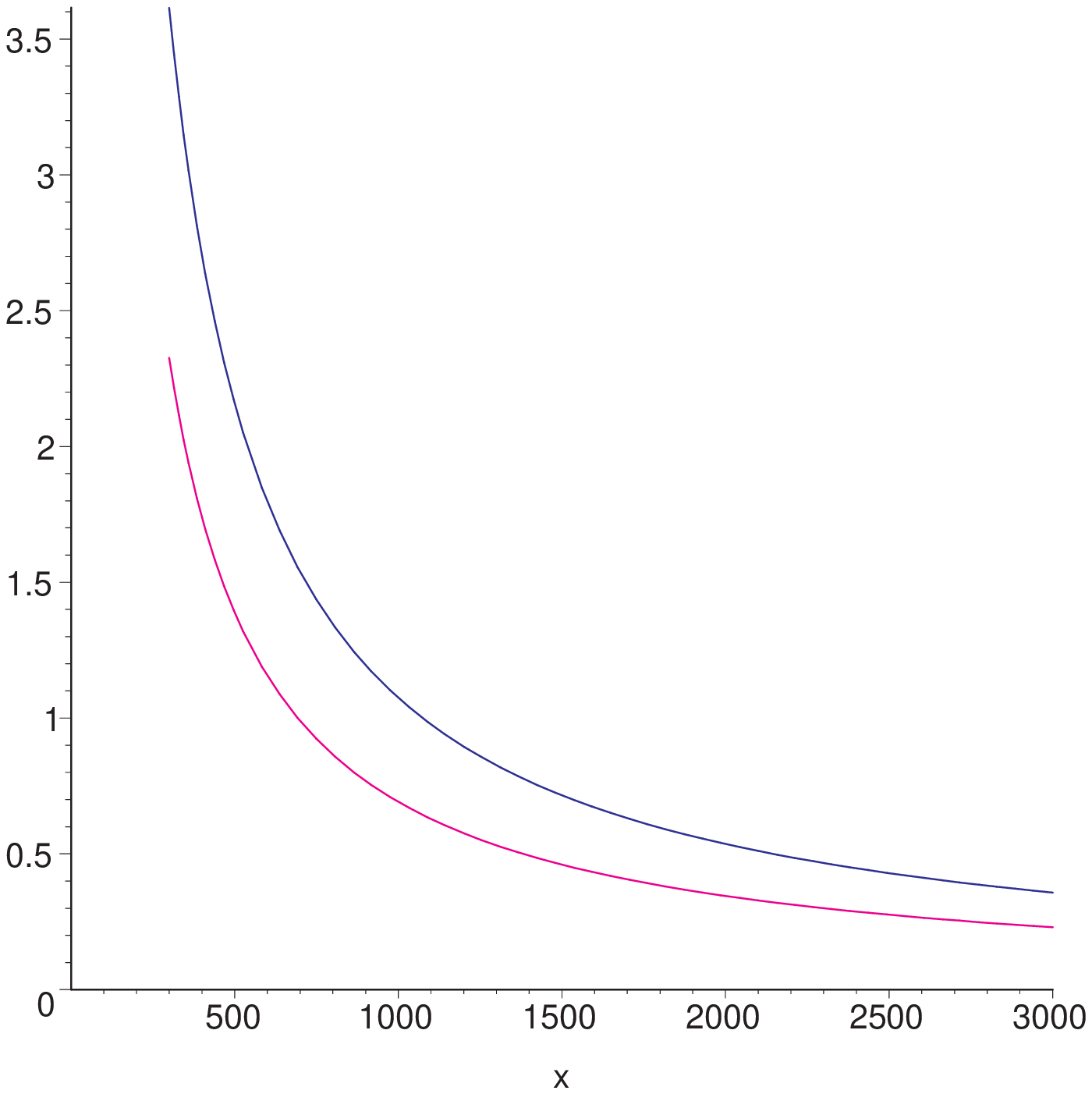}}
\put(-30,-83) {\includegraphics[height=5.6cm,width=5.6cm]{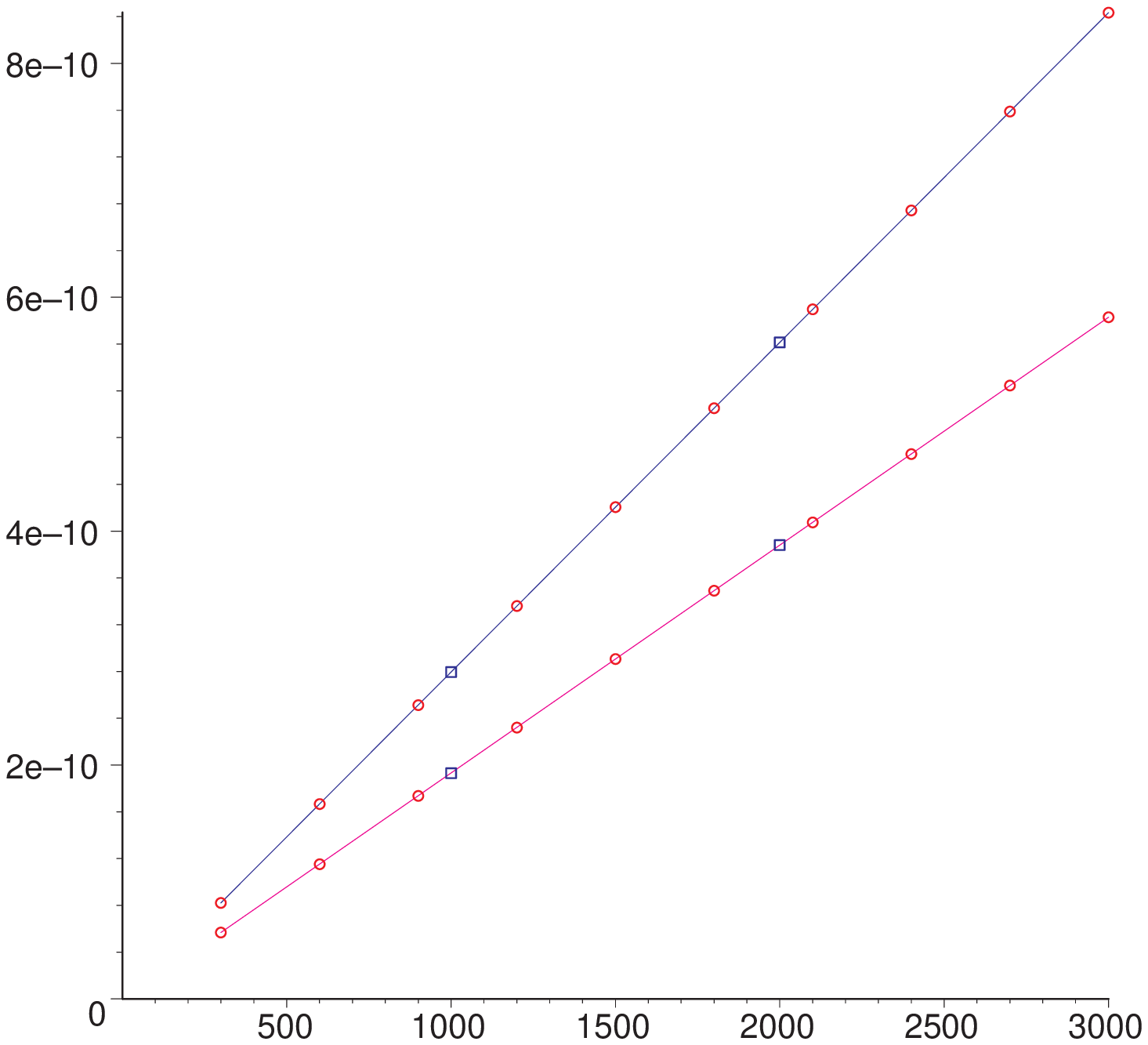}}
\put(40,-86)  {\includegraphics[height=5.6cm,width=5.6cm]{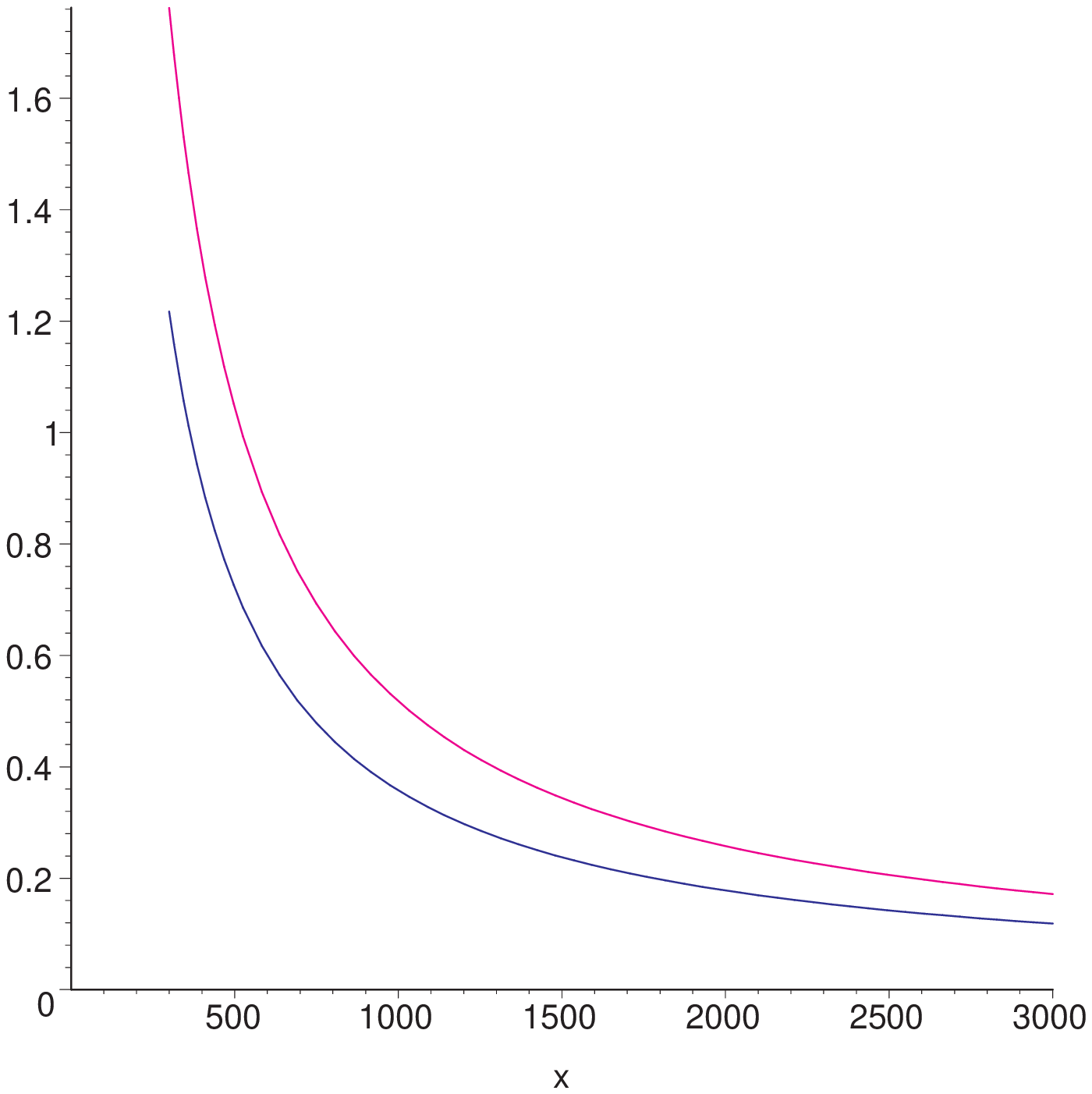}}

\put(19,-24){\scriptsize $n$}
\put(-25,24){\scriptsize $\mathfrak{p}_n$}
\put(8,20){\color{magenta}\tiny\texttt{CGCGC}}
\put(12,5){\color{blue}\tiny\texttt{AAAAA}}

\put(90,-24){\scriptsize $n$}
\put(45,24){\scriptsize $\Ex(T_n)/10^7$}
\put(60,-22){\color{magenta}\tiny\texttt{CGCGC}}
\put(80,-18){\color{blue}\tiny\texttt{AAAAA}}

\put(19,-79){\scriptsize $n$}
\put(-25,-30){\scriptsize $\mathfrak{p}_n$}
\put(0,-40){\color{blue}\tiny\texttt{CCCCCCCCCC}}
\put(12,-55){\color{magenta}\tiny\texttt{ATATATATAT}}

\put(90,-79){\scriptsize $n$}
\put(45,-32){\scriptsize $\Ex(T_n)/10^{10}$}
\put(56,-77){\color{blue}\tiny\texttt{CCCCCCCCCC}}
\put(80,-73){\color{magenta}\tiny\texttt{ATATATATAT}}



\end{picture}
\ \vspace*{8cm}
\caption{\label{fig:flexn} {\bf Plots of the probability $\mathfrak{p}_n$ (left)
  and of the expected waiting time $\Ex(T_n)$ (right)}.
(Top) $b=\texttt{AAAAA}$ (blue) and $b'=\texttt{CGCGC}$ (magenta);
(Down) $b=\texttt{CCCCCCCCCC}$ (blue) and $b'=\texttt{ATATATATAT}$ (magenta).
In the linear plots of the probability, the anchors values for $n=1000$
and $n=2000$ (computed by automata) are represented by boxes; the straight lines are the
straight lines going through the corresponding points and the circles
are
test values also computed by automata. The fit is perfect as expected
from singularity analysis.
}
\end{figure}
In Section~\ref{sec:auto} we considered by automata a parallel
computation on two sequences, $S(0)$ and $S(1)$.

It is possible to do a relevant
mathematical analysis with the 
random sequence $S(0)$ only. The corresponding computations have
however
a much higher complexity than the automaton approach.
This analysis is defined on counting in a random sequence $S(0)$
the number of putative-hit positions where, 
given a $k$-mer $b$, a putative-hit position is any position of $S(0)$
that
can lead by mutation to an occurrence of $b$ is $S(1)$, 
assuming that a single mutation has occurred.

For any $k$-mer $b$ \cite{Nicodeme2011}  provides
a combinatorial
construction  using clumps (see~\cite{BaClFaNi08})
that (i) considers {\sl all} the sequences that avoid the $k$-mer $b$,
and (ii)  counts {\sl all} the {\sl putative-hit} position
in these sequences.

In the following, let $H_n$ denote the number of putative-hit positions
in a sequence $S(0)$ randomly chosen within the set of sequences  
of length $n$ that do not contain the $k$-mer
$b$, where the letters are drawn with respect to the distribution
$\nu$
and where we put a probability mass $1$ to the set~\footnote{This is
  done by unconditioning with respect to the fact that $b$ does not
  occur in $S(0)$,  {\it i.e} by dividing the resulting expressions by
 $\Pr(S(0)\not\in \mathcal{A}^{\star}b\mathcal{A}^{\star}$);
see Equation \eqref{eq:S1S0}.}. As a consequence of  singularity
analysis
of rational functions, \cite{Nicodeme2011} 
proves  
that
\begin{equation}
\label{eq:ExpHn}
\Ex(H_n)=c_1\!\times\! n +c_2 +O(A^n)\qquad (A<1).
\end{equation}
It is clear that, using the asymptotic  Landau's
$\Theta$ notation, we do not have
\[\mathfrak{p}_n = \Theta (\Ex(H_n)),\]
since, for $n$  large enough, this would imply that $\mathfrak{p}_n>1$.
However, for 
\[\max_{\alpha\neq
  \beta\in\mathcal{A}}(p_{\alpha,\beta}(1))\ll
1\quad\text{and}\quad n \ll
1\big/\!\!\max_{\alpha\neq\beta\in\mathcal{A}}(p_{\alpha,\beta}(1)),\]
the probability that
two or more putative-hit positions simultaneously mutate to provide the
$k$-mer $b$ in sequence $S(1)$ is an event of second order small
probability.
With these conditions, we have
\begin{equation}
\label{eq:asymppn}
\mathfrak{p}_n \approx \rho_{b,\nu,p}\times \Ex(H_n)=  \rho_{b,\nu,p}\times (c_1\!\times\! n
+c_2) +O(A^n),
\end{equation}
where 
$\rho_{b,\nu,p}$ is a constant of the order of magnitude of the
constants
$p_{\alpha, \beta}(1)$ with $\alpha\neq \beta$, its value depending upon
these constants, the distribution $\nu$ and the correlation structure of
the $k$-mer $b$. See Figure~\ref{fig:flexn} for examples.

\paragraph{Available data.}\footnotesize \hskip-2ex  URL
 \scriptsize\mbox{\url{http://www.lix.polytechnique.fr/Labo/Pierre.Nicodeme/BNN/Waitforkmers.tar.gz}}
\footnotesize
provides access to the values of the expected waiting time $\Ex(T_n)$ and the
probability $\mathfrak{P}_n$ for $n=1000$ and $n=2000$ for all
$k$-mers with $k$ from $5$ to $10$. It is
therefore possible to compute $\mathfrak{p}_n$ and $\Ex(T_n)$ for all these $k$-mers for all
$n$ from these data. It took 10 hours to compute the data.
\normalsize
\section{Conclusion}
\label{sec:conclusion}
Using automata theory, we have developed a new procedure to compute the waiting time until a given TF binding site emerges at random in a human promoter sequence. In contrast to \cite{BehVin2010}, we do not have to rely on any assumptions regarding the overlap structure of the TF binding site of interest. Thus, our computations are more accurate. Assuming model M0, whose parameters have been estimated in the same way as in \cite{{BehVin2010}}, applying our automaton approach to all $k$-mers, $k$ ranging from 5 to 10, and comparing the resulting expected waiting times to those obtained by \cite{BehVin2010}, we particularly observe that highly autocorrelated words like \texttt{CCCCC} or \texttt{AAAGG} actually tend to emerge slower than predicted by \cite{BehVin2010}. This slowdown can attain up to 40\%, e.g. according to \cite{BehVin2010}, \texttt{CCCCC} is predicted to be created in a human promoter of length 1~kb in around 6.304~Mgen while our more accurate method predicts it be  generated in around 9.105~Mgen. 
We have shown that existing TF binding sites (from the database
JASPAR; \cite{jaspar}) feature a significant proportion of
autocorrelation. Therefore the assumption of \cite{BehVin2010} that TF
binding sites do not appear self-overlapping when computing waiting
times is problematic. The new automaton approach now incorporates the
possibility of TF binding sites appearing self-overlapping into the
model. Hence, the automaton approach highly improves the accuracy of
the estimations for waiting times. We observed a linear behaviour with
respect to the length of the promoters for
the probability of finding a $k$-mer at generation $1$ that is not present at
generation $0$. This implies a highly flexible and efficient approach
for computing
this probability for any promoter length, and in particular for lengths
of highest interest, i.e. between 300 and 3000~bp. This also induces a
hyperbolic behaviour for the waiting time.

\paragraph{Acknowledgements.} We thank Martin Vingron who initiated
the previous work of \cite{BehVin2010}, of which the present article is a
follow-up.

\paragraph{Disclosure statement.} No competing financial interests exist.


\begin{thebibliography}{21}
\providecommand{\natexlab}[1]{#1}
\providecommand{\url}[1]{\texttt{#1}}
\providecommand{\urlprefix}{URL }

\bibitem[{Arndt and Hwa(2005)}]{arndt2}
Arndt, P.~F. and Hwa, T., 2005.
\newblock Identification and measurement of neighbor-dependent nucleotide
  substitution processes.
\newblock \emph{Bioinformatics} 21, 2322--2328.

\bibitem[{Bassino \emph{et~al.}(2008)Bassino, Cl\'ement, Fayolle, and
  Nicod\`eme}]{BaClFaNi08}
Bassino, F., Cl\'ement, J., Fayolle, J., and Nicod\`eme, P., 2008.
\newblock Constructions for clump statistics.
\newblock In Jacquet, P., ed., \emph{Proceedings of the Fifth Colloquium on
  Mathematics and Computer Science, Blaubeuren, Germany}, 183--198. DMTCS.
\newblock \
  \\\noindent\mbox{\small\url{http://www-lipn.univ-paris13.fr/~bassino/publica%
tions/mathinfo08.pdf}}.

\bibitem[{Behrens and Vingron(2010)}]{BehVin2010}
Behrens, S. and Vingron, M., 2010.
\newblock Studying the evolution of promoters: a waiting time problem.
\newblock \emph{J. Comput. Biol} 17, 1591--1606.
\newblock \
  \\\noindent\mbox{\small\url{http://www.liebertonline.com/doi/full/10.1089/cm%
b.2010.0084}}.

\bibitem[{Crochemore and Rytter(1994)}]{CroRyt94}
Crochemore, M. and Rytter, W., 1994.
\newblock \emph{Text Algorithms}.
\newblock Oxford University Press.

\bibitem[{Dowell(2010)}]{dowell}
Dowell, R.~D., 2010.
\newblock Transcription factor binding variation in the evolution of gene
  regulation.
\newblock \emph{Trends in Genetics} 26, 468 -- 475.

\bibitem[{Duret and Arndt(2008)}]{arndt3}
Duret, L. and Arndt, P.~F., 2008.
\newblock The impact of recombination on nucleotide substitutions in the human
  genome.
\newblock \emph{PLoS Genet.} 4.

\bibitem[{Durrett and Schmidt(2007)}]{durrett}
Durrett, R. and Schmidt, D., 2007.
\newblock Waiting for regulatory sequences to appear.
\newblock \emph{Ann. Appl. Probab.} 17, 1--32.

\bibitem[{Flajolet and Sedgewick(2009)}]{FlajoletSedgewick2009}
Flajolet, P. and Sedgewick, R., 2009.
\newblock \emph{Analytic Combinatorics}.
\newblock Cambridge University Press.

\bibitem[{Goulden and Jackson(1983)}]{GoJa83}
Goulden, I. and Jackson, D., 1983.
\newblock \emph{Combinatorial {E}numeration}.
\newblock John Wiley.
\newblock New-York.

\bibitem[{Guibas and Odlyzko(1981{\natexlab{a}})}]{GuiOdl81a}
Guibas, L. and Odlyzko, A., 1981{\natexlab{a}}.
\newblock Periods in strings.
\newblock \emph{J. Combin. Theory} A, 19--42.

\bibitem[{Guibas and Odlyzko(1981{\natexlab{b}})}]{GuiOdl81b}
Guibas, L. and Odlyzko, A., 1981{\natexlab{b}}.
\newblock Strings overlaps, pattern matching, and non-transitive games.
\newblock \emph{J. Combin. Theory} A, 108--203.

\bibitem[{Hopcroft \emph{et~al.}(2001)Hopcroft, Motwani, and Ullman}]{HU01}
Hopcroft, J., Motwani, R., and Ullman, J., 2001.
\newblock \emph{Introduction to Automata Theory, Languages and Computation}.
\newblock Addison-Wesley.

\bibitem[{Karlin and Taylor(1975)}]{KarTay75}
Karlin, S. and Taylor, H., 1975.
\newblock \emph{A First Course in Stochastic Processes}.
\newblock Academic Press.
\newblock Second Edition, 557 pages.

\bibitem[{Kunarso \emph{et~al.}(2010)Kunarso, Chia, Jeyakani, Hwang, Lu, Chan,
  Ng, and Bourque}]{kunarso}
Kunarso, G., Chia, N.-Y., Jeyakani, J., Hwang, C., Lu, X., Chan, Y.-S., Ng,
  H.-H., and Bourque, G., 2010.
\newblock Transposable elements have rewired the core regulatory network of
  human embryonic stem cells.
\newblock \emph{Nature Genetics} 42, 631--634.

\bibitem[{Lothaire(2005)}]{Lot05}
Lothaire, M., 2005.
\newblock \emph{Applied Combinatorics on Words}.
\newblock Encyclopedia of Mathematics. Cambridge University Press.

\bibitem[{Nicod\`eme(2011)}]{Nicodeme2011}
Nicod\`eme, P., 2011.
\newblock A clump analysis for waiting times in {DNA} evolution.
\newblock Personal communication, \ \\
  \small\url{http://www.lix.polytechnique.fr/Labo/Pierre.Nicodeme/pncpm12.pdf}.

\bibitem[{Odom \emph{et~al.}(2007)Odom, Dowell, Jacobsen, Gordon, Danford,
  MacIsaac, Rolfe, Conboy, Gifford, and Fraenkel}]{odom}
Odom, D.~T., Dowell, R.~D., Jacobsen, E.~S., Gordon, W., Danford, T.~W.,
  MacIsaac, K.~D., Rolfe, P.~A., Conboy, C.~M., Gifford, D.~K., and Fraenkel,
  E., 2007.
\newblock Tissue-specific transcriptional regulation has diverged significantly
  between human and mouse.
\newblock \emph{Nat. Genet.} 39, 730--732.

\bibitem[{Portales-Casamar \emph{et~al.}(2010)Portales-Casamar, Thongjuea,
  Kwon, Arenillas, Zhao, Valen, Yusuf, Lenhard, Wasserman, and
  Sandelin}]{jaspar}
Portales-Casamar, E., Thongjuea, S., Kwon, A.~T., Arenillas, D., Zhao, X.,
  Valen, E., Yusuf, D., Lenhard, B., Wasserman, W.~W., and Sandelin, A., 2010.
\newblock {JASPAR 2010}: the greatly expanded open-access database of
  transcription factor binding profiles.
\newblock \emph{Nucl. Acids Res.} 38, D105--110.

\bibitem[{Schmidt \emph{et~al.}(2010)Schmidt, Wilson, Ballester, Schwalie,
  Brown, Marshall, Kutter, Watt, Martinez-Jimenez, Mackay, Talianidis, Flicek,
  and Odom}]{schmidt}
Schmidt, D., Wilson, M.~D., Ballester, B., Schwalie, P.~C., Brown, G.~D.,
  Marshall, A., Kutter, C., Watt, S., Martinez-Jimenez, C.~P., Mackay, S.,
  Talianidis, I., Flicek, P., and Odom, D.~T., 2010.
\newblock Five-vertebrate chip-seq reveals the evolutionary dynamics of
  transcription factor binding.
\newblock \emph{Science} 328, 1036--1040.

\bibitem[{Stone and Wray(2001)}]{stone}
Stone, J.~R. and Wray, G.~A., 2001.
\newblock Rapid evolution of cis-regulatory sequences via local point
  mutations.
\newblock \emph{Mol. Biol. Evol.} 18, 1764--1770.

\bibitem[{Wray \emph{et~al.}(2003)Wray, Hahn, Abouheif, Balhoff, Pizer,
  Rockman, and Romano}]{wray}
Wray, G.~A., Hahn, M.~W., Abouheif, E., Balhoff, J.~P., Pizer, M., Rockman,
  M.~V., and Romano, L.~A., 2003.
\newblock The evolution of transcriptional regulation in eukaryotes.
\newblock \emph{Mol. Biol. Evol.} 20, 1377--1419.

\end{thebibliography}

\end{document}